\documentclass{elsart}
\usepackage{graphicx}
\usepackage{amssymb,amsmath}
\journal{Chemical Physics Letters}

\begin{document}

\begin{frontmatter}
\title{Adsorption preference reversal phenomenon from multisite-occupancy theory for two-dimensional lattices}

\author{D. A. Matoz-Fernandez and}
\author{A. J. Ramirez-Pastor\corauthref{cor1}}

\address{Departamento de F\'{\i}sica, Instituto de F\'{\i}sica
Aplicada, Universidad Nacional de San Luis-CONICET, Ej\'ercito de
los Andes 950, D5700BWS, San  Luis, Argentina}

\thanks[cor1]{Corresponding author. E-mail: antorami@unsl.edu.ar}


\begin{abstract}

The statistical thermodynamics of polyatomic species mixtures adsorbed on two-dimensional substrates was developed on a generalization in the spirit of the lattice-gas model and the
classical Guggenheim-DiMarzio approximation. In this scheme, the coverage and temperature dependence of the Helmholtz free energy and chemical potential are given. The formalism leads to the exact statistical thermodynamics of binary mixtures adsorbed in one dimension, provides a close approximation for two-dimensional systems accounting multisite occupancy and allows to discuss the dimensionality and lattice structure effects on the known phenomenon of adsorption preference reversal.

\end{abstract}

\end{frontmatter}

\section{Introduction}
\label{intro}

Two previous articles \cite{JCP8,CPL1} referred to as papers I and II, respectively, were devoted to the study of binary mixtures adsorption. In paper I, the multicomponent adsorption of polyatomic species was described as a
fractional statistics problem, based on Haldane's statistics \cite{Haldane,Wu}. The thermodynamic functions calculated for a monomer-dimer mixture were applied to describe the adsorption of methane-ethane mixtures in zeolites. The obtained results revealed that, at low pressure, the adsorbed phase is almost entirely ethane, but at high-pressure methane displaces ethane reproducing the adsorption preference reversal (APR) phenomenon observed in Monte Carlo simulations, mean-field theory, and exact calculations \cite{Du,Khettar,Ayache,Dunne} of the system under study.

In Refs. \cite{Ayache,Dunne}, the authors had shown how the competition between two species in presence of repulsive mutual interactions can be responsible of the displacement of one species by the other.
The results in paper I contributed to the understanding of the phenomenon, showing that if realistic single species adsorption energy values are used when studying mixtures, the APR will appear as a result of the difference of size (or number of occupied sites) between the adsorbed species. Thus, to introduce repulsive lateral interactions in the adsorbate, as done in Refs. \cite{Ayache,Dunne}, is a fictitious or effective way of taking into account geometric or steric effects by means of energetic arguments. A real description of the displacement of flat ethane by methane and upright ethane should include multisite-occupancy adsorption.

Paper II was a step further, addressing the rigorous statistical thermodynamics of $s$-mer (particle occupying $s$ lattice sites)–$k$-mer (particle occupying $k$ lattice sites) mixtures adsorbed on one-dimensional substrates. The formalism presented in paper II was the first exact model of adsorption of polyatomic mixtures in zeolites and allows to demonstrate that the APR phenomenon is the result of the difference of size (or number of occupied sites) between the adsorbed species. In other words, the results in paper II revealed that a real description of the phenomenon of APR may be severely misunderstood, if the polyatomic character of the adsorbate is not properly incorporated in the thermodynamic functions from which experiments are interpreted.

In contrast to the statistics for simple particles, where the arrangement of the adsorption sites in space is immaterial, the structure of lattice space plays such a
fundamental role in determining the statistics of $k$-mers. Then, it is of interest and of value to inquire how a specific lattice structure influences the main thermodynamic properties
of adsorbed polyatomics. In this sense, the aim of the present work is to extend the study in paper II to two-dimensional lattices. The problem is not only of theoretical interest, but also has practical importance
(most of the experiments in surface science are carried out in two dimensions).

\section{Theory: lattice model for adsorption of alkane binary mixtures}

In this paper, the adsorption of a binary mixture of straight
rigid $k$-mers and straight rigid $l$-mers on two-dimensional
lattices is considered. The $k$($l$)-mers are assumed to be
composed by $k$($l$) identical units in a linear array with
constant bond length equal to the lattice constant $a$. Without
any loss of generality, we assume that $l < k$. The substrate is
modeled by a two-dimensional array of $M$ sites ($M \rightarrow
\infty$) and connectivity $\gamma$, where periodic boundary
conditions apply. Under this condition, all lattice sites are
equivalent, hence border effects will not enter our derivation.
The $k$($l$)-mers can only adsorb flat on the surface occupying
$k$($l$) contiguous lattice sites. In addition, double site
occupancy is not allowed as to represent properties in the
monolayer regime. Since different particles do not interact with
each other, all configurations of $N_k$ $k$-mers and $N_l$
$l$-mers on $M$ sites are equally probable. Then, the canonical
partition function $Q(M,N_k,N_l,T)$ equals the total number of
configurations, $\Omega (M,N_k,N_l)$, times a Boltzmann factor
including the total interaction energy between adparticles and
substrate $\epsilon(N_k,N_l)$,
\begin{equation}
Q(M,N_k,N_l,T) = \Omega(M,N_k,N_l)\exp \left[-\frac{\epsilon(N_k,N_l)}{ k_BT}\right].
\end{equation}

In order to calculate $\Omega (M,N_k,N_l)$, the DiMarzio's lattice theory can be used \cite{DiMarzio}. Let's start calculating the number of distinct ways to pack $N_k$ rigid rods onto a lattice with $d$
allowed orientations (directions),
\begin{equation}
\label{DiMarzio_k}
\begin{split}
\Omega_{k}(M,\lbrace N_{k} \rbrace_{d}) = & \frac{\prod_{i=1}^{d} \left[M- \left( k-1 \right)N_{k,i}
\right]!} {\left(M-k\,N_k \right)! \left( M! \right)^{d-1} \prod_{i=1}^{d} \left( N_{k,i} \right)!},
\end{split}
\end{equation}
where $N_{k,i}$ is the number of $k$-mers lying in the direction $i$ and $N_k=\sum_{i=1}^{d} N_{k,i}$ is the total number of \emph{k}-mers on the surface.

Now, using the DiMarzio counting scheme, the number of ways to place the $(j_1+1)$th $l$ type molecule onto the lattice (the subscript reminds us that we are discussing the orientation 1), given that $j_1$ $l$ molecules have already been placed in the direction 1 and $N_{k}$ type $k$ molecules have already been placed, is seen to be \cite{Peterson},
\begin{equation}
\nu_{j_1+1}=\left(M-k\,N_{k}-l\,j_1 \right)\left[\frac{M-k\,N_{k}-l\,j_1}{M-(k-1)N_{k,{1}}-(l-1)j_1} \right]^{l-1}.
\end{equation}
The total number of ways to place $N_{l,1}$ indistinguishable molecules onto the lattice in this orientation is,
\begin{equation}
\frac{\prod_{j_1=0}^{N_1-1}\nu_{j_1+1}}{\left( N_{l,1} \right)!}=
\frac{(M-k\,N_{k})!\left[M-(k-1)N_{k,1}-(l-1)N_{l,1}\right]!}{(M-k\,N_{k}-l\,N_{l,1})!\left[M-(k-1)N_{k,1}\right]!N_{l,1}!}.
\end{equation}
Similar expressions can be obtained for the other orientations and total numbers of ways to place $N_l$ hard rod molecules type $l$ when $N_{k}$ type $k$ have been place in the surface is,
\begin{equation}
\label{DiMarzio_l}
\begin{split}
\Omega_{l}(M,\lbrace N_{l} \rbrace_{d}) = & \frac{\left(M-k\,N_k \right)!\prod_{i=1}^{d} \left[M- \left( k-1 \right)N_{k,i}
- \left(l-1 \right)N_{l,i}\right]!} {\left(M-k\,N_k-l\,N_l \right)! \prod_{i=1}^{d} \left[M- \left( k-1 \right)N_{k,i}
\right]! \left( N_{l,i} \right)!}.
\end{split}
\end{equation}
Them the product obtained from Eqs. (\ref{DiMarzio_k}) and (\ref{DiMarzio_l}) gives the total number of ways to pack the molecules in the mixture:
\begin{equation}
\label{DiMarzio_lk}
\Omega(M,\lbrace N_{k} \rbrace_{d},\lbrace N_{l} \rbrace_{d})= \frac{\prod_{i=1}^{d}\left[M-\left(k-1\right)N_{k,i}-\left(l-1\right)N_{l,i}\right]!}{\left(M-k\,N_k-l\,N_l \right)! \left( M! \right)^{d-1} \prod_{i=1}^{d} \left( N_{k,i} \right)!\left( N_{l,i} \right)!}.
\end{equation}
Equation (\ref{DiMarzio_lk}) is exact when all  molecules live in one direction  \cite{CPL1}. For the case of an isotropic distribution of molecules, i.e., $N_{k(l),i}=(2/\gamma)N_{k(l)}$, then the appropriate generalization of Eq. (\ref{DiMarzio_lk}) is,
\begin{equation}
\label{Omega_g}
\Omega(M,N_k,N_l)=\frac{M!}{(M-kN_k-lN_l)!}\left\lbrace \frac{\left[M-(k-1)\frac{\gamma}{2}N_k-(l-1)\frac{\gamma}{2}N_l  \right]!}{M!\,N_k!\,N_l!} \right\rbrace^{\frac{\gamma}{2}}.
\end{equation}

In the canonical ensemble, the Helmholtz free energy $F(M,N_{k},N_{l},T)$ relates to $\Omega(M,N_k,N_l)$ through,
\begin{equation}
\label{FLibre}
\begin{split}
\beta F(M,N_k,N_l,T) &= -\ln Q(M,N_k,N_l,T) \\
& = -\ln \Omega(M,N_k,N_l)+ \beta \epsilon (N_k,N_l),
\end{split}
\end{equation}
where $\beta=1/k_BT$, $\epsilon (N_k,N_l)=\epsilon_k\,N_k+\epsilon_l\,N_l$ and $\epsilon_i$ represents the interaction energy of a $i$-mer ($i = k,l$) adsorbed on the substrate.

The chemical potential of the adsorbed species $i$, $\mu_{i,ads}$, can be calculated as \cite{Hill},
\begin{equation}
\label{MUi}
\mu_{i,ads}=\left({\frac{\partial F}{\partial N_i}}
\right)_{{N_j}} \  \  \  \  \  \  \  \  \    \{i,j=k,l\}.
\end{equation}
From Eqs. (\ref{Omega_g}-\ref{MUi}) it follows that,
\begin{equation}
\label{MUsups_k}
\begin{split}
\mu_{k,ads}=&(k-1)\ln\left[\frac{\gamma}{2}-\frac{(k-1)}{k}\theta_k-\frac{(l-1)}{l}\theta_l\right]+\ln\left(\frac{\theta_k}{k}\right)\\
&-k\ln\left(1-\theta_k-\theta_l\right)-k\ln\left(\frac{\gamma}{2}\right)+\beta\,\epsilon_k,
\end{split}
\end{equation}
and
\begin{equation}
\label{MUsups_l}
\begin{split}
\mu_{l,ads}=&(l-1)\ln\left[\frac{\gamma}{2}-\frac{(k-1)}{k}\theta_k-\frac{(l-1)}{l}\theta_l\right]+\ln\left(\frac{\theta_l}{l}\right)\\
&-l\ln\left(1-\theta_k-\theta_l\right)-l\ln\left(\frac{\gamma}{2}\right)+\beta\,\epsilon_l,
\end{split}
\end{equation}
where $\theta_i=iN_i/M$ represents the partial coverage of the species $i$ $\{i=k,l\}$. At equilibrium, the chemical potential of the adsorbed and gas
phase are equal. Then,
\begin{equation}\label{equis}
\mu_{k,ads}=\mu_{k,gas},
\end{equation}
and
\begin{equation}\label{equik}
\mu_{l,ads}=\mu_{l,gas},
\end{equation}
where $\mu_{k,gas}$ ($\mu_{l,gas}$) corresponds to $k$-mers ($l$-mers) in gas phase.

The chemical potential of each kind of molecule in an ideal gas mixture, at temperature $T$ and pressure $P$, is
\begin{equation}\label{mugask}
\beta \mu_{k,gas}=\beta \mu_k^0+\ln X P,
\end{equation}
and
\begin{equation}\label{mugasl}
\beta \mu_{l,gas}=\beta \mu_l^0+\ln (1-X)P,
\end{equation}
where $\mu_{k}^0$ and $\mu_{l}^0$ ($X$ and $1-X$) are the standard chemical potentials
(mole fractions) of $k$-mers and $l$-mers, respectively. In addition,
\begin{equation}
\beta \mu_i^0=-\ln \left[\left({2\pi m_ik_BT}\over
h^2\right)^{3/2}k_BT \right]\  \  \  \  \  \  \  \  \ \{i=k,l\}.
\label{mu00}
\end{equation}

\section{Adsorption preference reversal phenomenon in two dimensions}

Our model can be used for predict the behavior of APR phenomenon for $n$-alkanes mixtures.
As in Refs.\cite{CPL1,Ayache}, we start defining the $\Phi_i$'s
parameters,
\begin{equation}\label{Phikk}
\beta \Phi_k \equiv \beta \epsilon_{k}-\beta\mu_k^0-\ln X P,
\end{equation}
and
\begin{equation}\label{Phill}
\beta \Phi_l \equiv \beta \epsilon_{l}-\beta\mu_l^0-\ln (1-X)P.
\end{equation}
Then, from Eqs. (\ref{MUsups_k}-\ref{Phill}), it results that
\begin{equation}
\label{Phi_k}
\begin{split}
&(k-1)\ln\left[\frac{\gamma}{2}-\frac{(k-1)}{k}\theta_k-\frac{(l-1)}{l}\theta_l\right]+\ln\left(\frac{\theta_k}{k}\right)\\
&-k\ln\left(1-\theta_k-\theta_l\right)-k\ln\left(\frac{\gamma}{2}\right)+\beta\Phi_k=0,
\end{split}
\end{equation}
and
\begin{equation}
\label{Phi_l}
\begin{split}
&(l-1)\ln\left[\frac{\gamma}{2}-\frac{(k-1)}{k}\theta_k-\frac{(l-1)}{l}\theta_l\right]+\ln\left(\frac{\theta_l}{l}\right)\\
&-l\ln\left(1-\theta_k-\theta_l\right)-l\ln\left(\frac{\gamma}{2}\right)+\beta\Phi_l=0.
\end{split}
\end{equation}

Now, following the line of Refs.\cite{CPL1,Ayache}, we define
the quantity $A \equiv \exp{[\beta(\Phi_k-\Phi_l)]}$, which is
obtained from the equilibrium Eqs. (\ref{Phi_k}) and
(\ref{Phi_l}),
\begin{equation}
\label{Apara}
\begin{split}
A&=\frac{1-X}{X}\exp{\left[\beta(\epsilon_k-\epsilon_l)-\beta(\mu_k^0-\mu_l^0)
\right]}
\\
&=\frac{k\theta_l}{l\theta_k}\frac{\left(1-\theta_k-\theta_l\right)^{k-l}}{\left[\frac{\gamma}{2}-\frac{(k-1)}{k}\theta_k-\frac{(l-1)}{l}\theta_l\right]^{k-l}}\left(\frac{\gamma}{2}\right)^{k-l}.
\end{split}
\end{equation}

As shown in Fig. 1, if the APR phenomenon occurs, then there
exists some value of coverage, $\theta^*$ ($0<\theta^*<1$), at
which the partial isotherms coincide ($\theta_k=\theta_l=\theta^*$). The crossing point $(\mu^*,\theta^*)$ separates two adsorption regimes. Thus, for $\mu < \mu^*$, $\theta_k$ is larger than $\theta_l$. This tendency is reverted for $\mu > \mu^*$, where $\theta_l$ is larger than $\theta_k$. Then, the existence of the point $\theta^*$ is directly related to the displacement of the species $k$ by the species $s$, and, consequently, to the presence of APR phenomenon. The value of $\theta^*$ can be
obtained from Eq. (\ref{Apara}),
\begin{equation}
\label{thetaAPR}
\theta^*=\frac{1-\left(\frac{lA}{k}\right)^{1/(k-l)}}{2-\left(\frac{k-1}{k}+\frac{l-1}{l}\right)\left(\frac{2}{\gamma}\right)\left(\frac{lA}{k}\right)^{1/(k-l)}},
\end{equation}
with the condition that $0<A<1$. This condition is satisfied given
that the adsorption energies corresponding to linear alkanes are
attractive and increase, in absolute value, linearly with the
chain length \cite{Chaer,Unge,Silva,LANG10,IECR1}.

In the following, the dependence of $\theta^*$ on the adsorption energies $\epsilon_k$ and $\epsilon_l$, the geometry $\gamma$ and the mole fraction $X$ will be investigated. As is
common in the literature \cite{LANG10,IECR1}, we adopt a ``bead segment" model of the alkane chains, in which each methyl (bead) group occupies one adsorption site on the surface with adsorption energy $\epsilon_0$. Under this consideration, $\epsilon_k=k \epsilon_0$ and $\epsilon_l=l \epsilon_0$. In addition, the values of $\beta \mu_k^0$ and $\beta \mu_l^0$ were obtained by using Eq. (\ref{mu00}) with $m_i$ equal to the molar mass of an alkane of size $i$, $C_iH_{(2i+2)}$ \cite{Lide}, and $T$=300 K.

Figure 2 shows the dependence of $\theta^*$ on $k$ for a $(l-k)$ mixture with $l=1$\footnote{The $(l-k)$ mixture with $l=1$ and $k \geq 2$ is the simplest case of an alkane binary mixture and contains
all the properties of the multisite occupancy adsorption.}, $\gamma=4$ (square lattice), $X=0.5$ and different values of the adsorption energy $\beta \epsilon_0(= -3, -6, -12, -24)$.
The curves increase monotonically with the size $k$ and converge asymptotically to a finite value as $k \rightarrow \infty$. This finite value tends to 0.5 for high values of $\beta \epsilon_0$. In the case of the figure, $\theta^*(\infty) \approx 0.375, 0.465, 0.497$, and $0.5$ for $\beta \epsilon_0= -3, -6, -12$, and $-24$, respectively.

In order to analyze the effect of lattice geometry, the coverage $\theta^*$ was calculated as a function of $k$ for a typical case ($\beta \epsilon_0=-3$, $l=1$ and $X=0.5$) and one-dimensional ($\gamma=2$), honeycomb ($\gamma=3$), square ($\gamma=4$) and triangular ($\gamma=6$) lattices. The results are shown in Fig.3. The general form of the curves is similar to that reported in Fig. 2. In this case, the limit value $\theta^*(\infty)$ strongly depends on the connectivity $\gamma$, being an increasing function of $\gamma$.

To conclude the analysis of Eq. (\ref{thetaAPR}), the dependence of $\theta^*$ on the molar fraction $X$ is analyzed in Fig. 4. The parameters chosen for this study were $\beta \epsilon_0=-3$, $l=1$ and $\gamma=4$. Several conclusions can be drawn from the figure.

For values of $X$ between $X=0[X \approx 0.448]$ and $X \approx 0.448[X=1]$ and $k$ varying between $k=0$ and $k \approx 20$, $\theta^*$ is a decreasing[increasing] function of the parameter $k$. For $k> 20$, all curves merge into one single curve which converges to a constant value as $k \rightarrow \infty$. This constant value (near 0.38 in the figure) depends on the lattice connectivity and the value of $A$.

For small values of $k$, APR phenomenon disappears as the molar fraction is above a certain critical value $X=X^*$. This behavior determines the phase diagram shown in the inset of Fig. 4, which does not change with
the lattice geometry, and only depends on the differences between the relative standard chemical potentials and the adsorption energies.

In summary, the statistical thermodynamics of polyatomic species mixtures adsorbed on two-dimensional substrates was developed on a generalization in the spirit of the lattice-gas model and the
classical Guggenheim-DiMarzio approximation. The theoretical formalism allows to study the problem of adsorption of alkane binary mixtures, leads to the exact solution in one dimension, provides a close approximation for two-dimensional systems accounting multisite occupancy and represents a contribution to the understanding of the phenomenon of adsorption preference reversal.

\section*{Acknowledgments}

This work was supported in part by CONICET (Argentina) under
project number PIP 112-201101-00615; Universidad Nacional de San
Luis (Argentina) under project 322000 and the National Agency of Scientific and
Technological Promotion (Argentina) under project  PICT-2010-1466.

\newpage

\section{Figure Captions}
\begin{itemize}

\item[Figure 1:] Typical partial adsorption isotherms for a $k$-mer-$l$-mer mixture in the presence of APR phenomenon. The crossing point $(\mu^*,\theta^*)$ separates two adsorption
regimes. Thus, for $\mu< \mu^*$, $\theta_k$ is larger than $\theta_l$. This tendency is reverted for $\mu > \mu^*$, where $\theta_l$ is larger than $\theta_k$. Then, the existence of the point $\theta^*$ is directly related to the displacement of the species $k$ by the species $l$, and, consequently, to the occurrence of APR phenomenon.

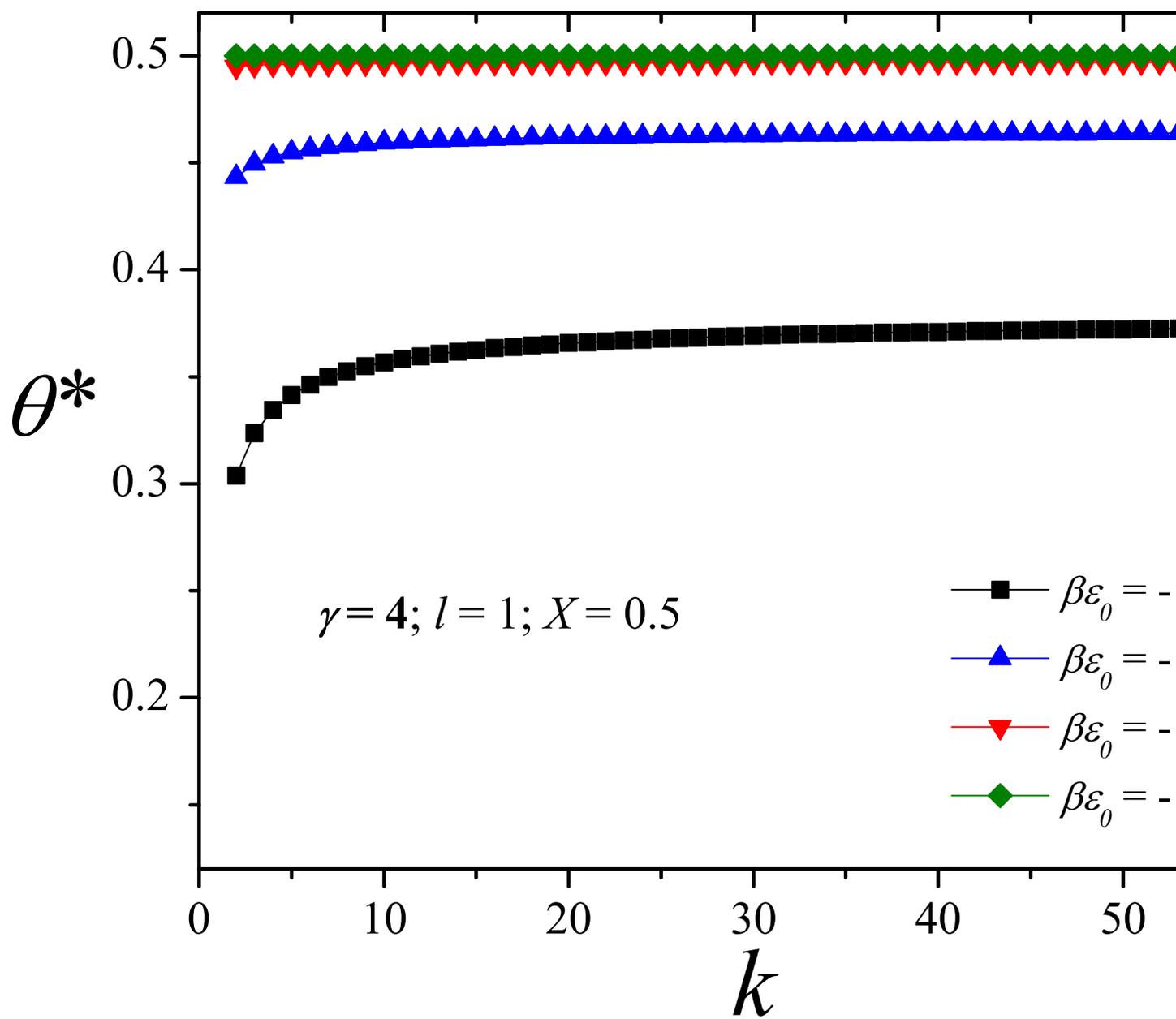
\item[Figure 2:] Dependence of $\theta^*$ on $k$ for a $(l-k)$ mixture with $l=1$, $X=0.5$ and different values of the adsorption energy $\beta \epsilon_0$ as indicated.

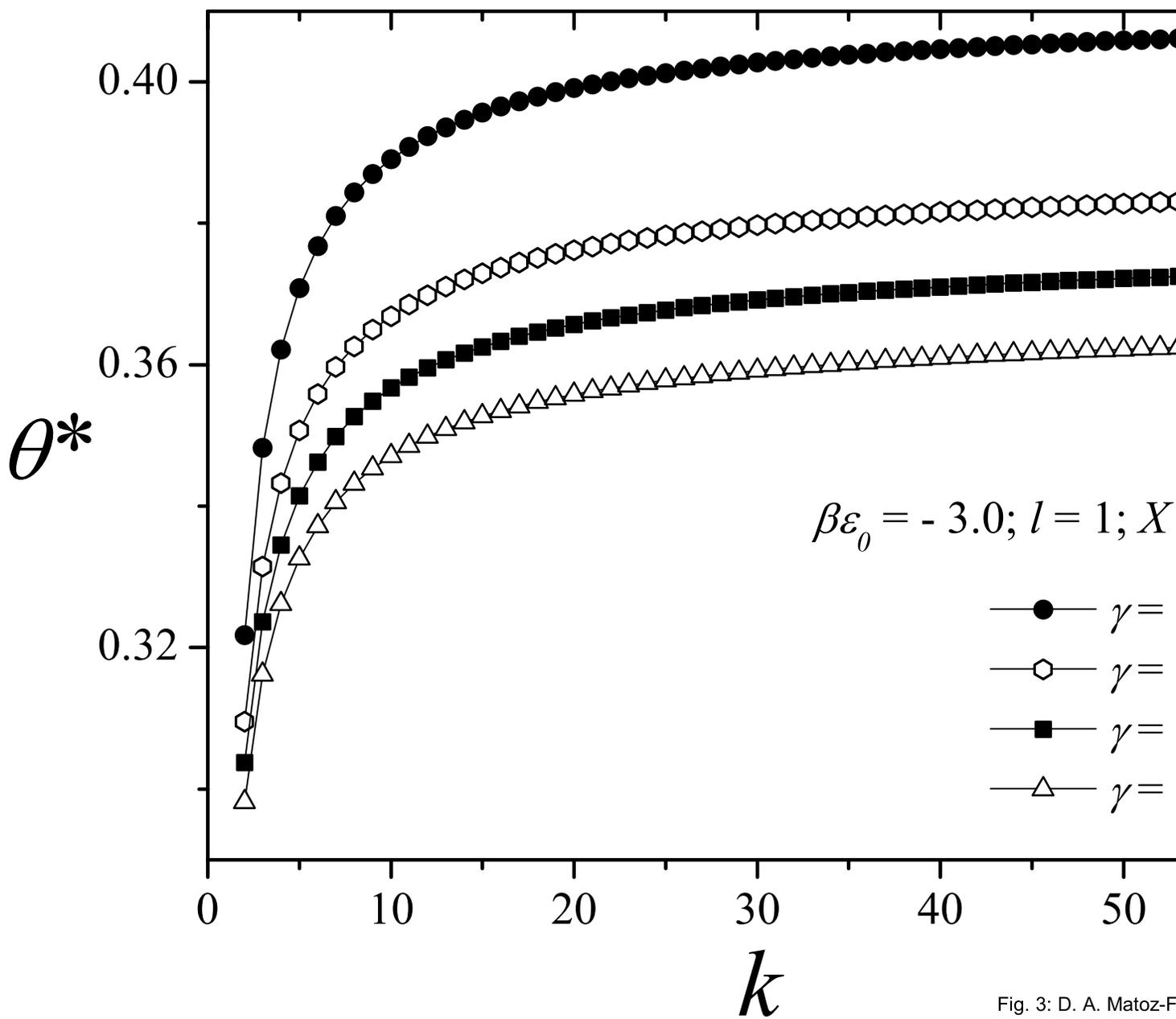
\item[Figure 3:] $\theta^*$ as a function of $k$ for a typical case ($\beta \epsilon_0=-3$, $l=1$ and $X=0.5$) and different geometries as indicated.

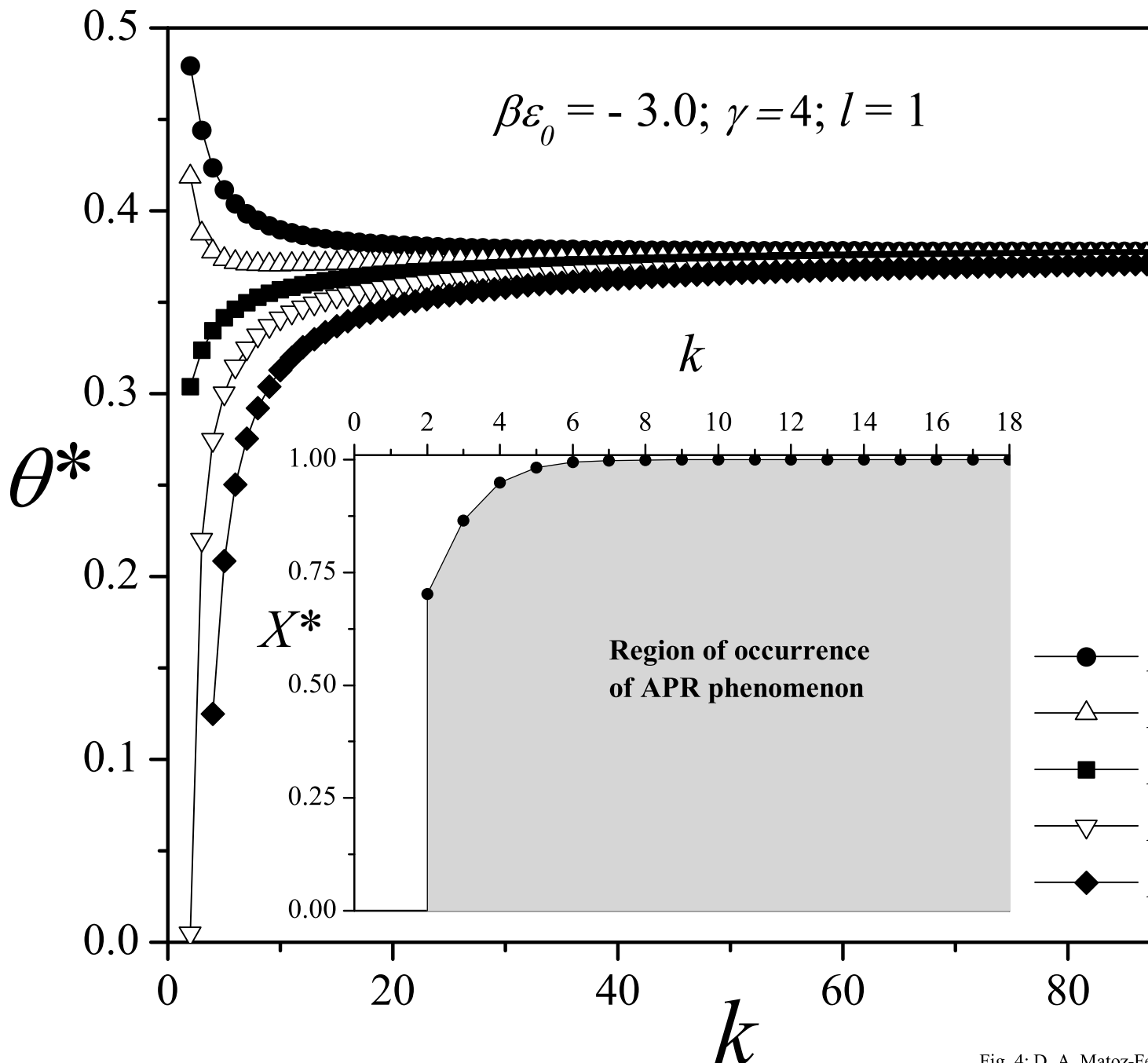
\item[Figure 4:] $\theta^*$ as a function of $k$ for a typical case ($\beta \epsilon_0=-3$, $l=1$ and $\gamma=4$) and different values of the molar fraction $X$ as indicated.

\end{itemize}

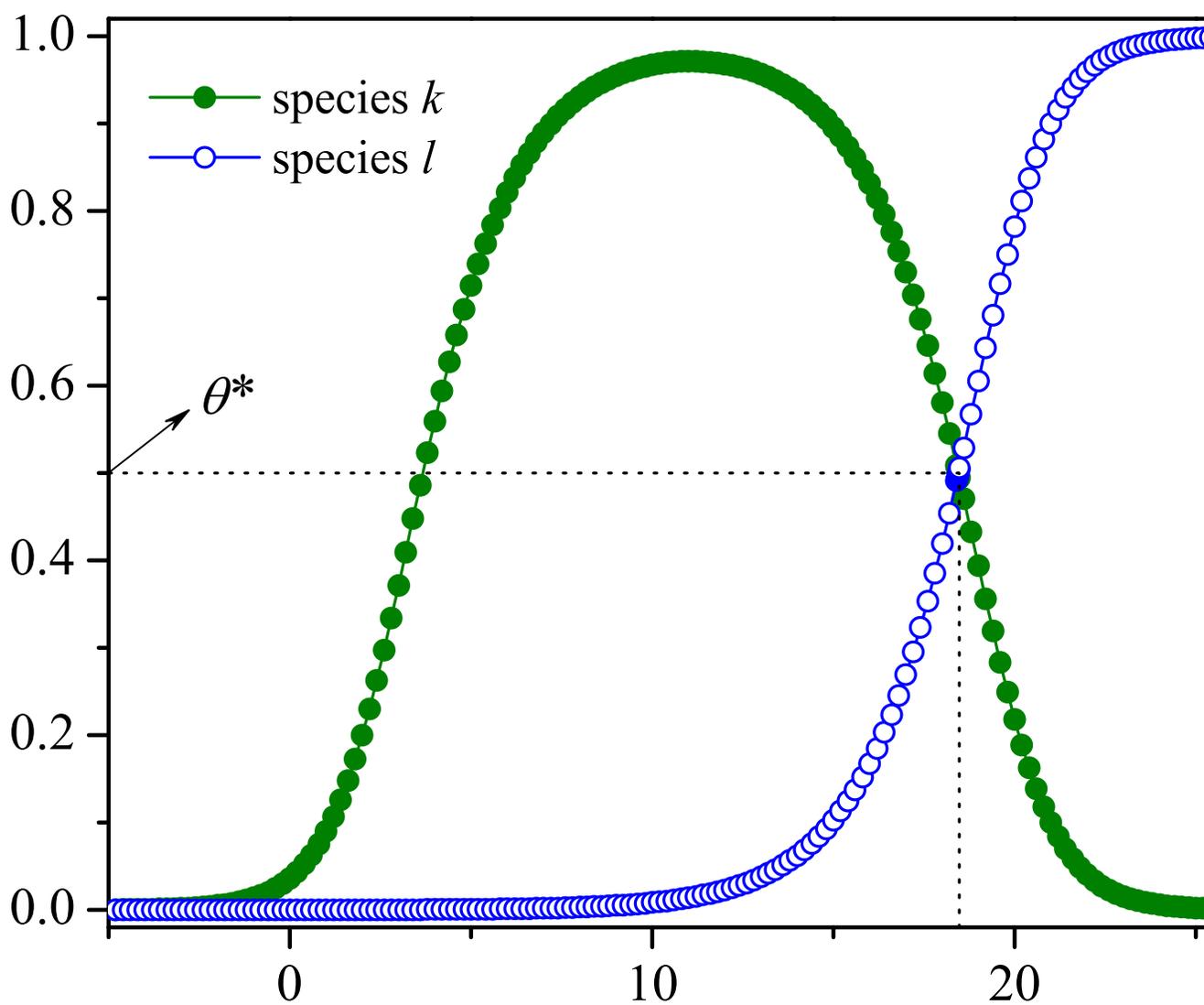
\end{document}